\documentstyle[aps,prb]{revtex} 
\input epsf.sty 

\begin{document}
\renewcommand{\theequation}{\arabic{section}.\arabic{equation}} 

\twocolumn[ 

\hsize\textwidth\columnwidth\hsize\csname@twocolumnfalse\endcsname 
\draft


\title {\bf Deconfinement of Spinons on Critical Points: \\
Multi-Flavor CP$^1$ + U(1) Lattice Gauge Theory in Three Dimensions
}

\author{Shunsuke Takashima and Ikuo Ichinose}  
\address{ Department of Applied Physics,
Nagoya Institute of Technology, Nagoya, 466-8555 Japan
}
\author{Tetsuo Matsui} 
\address{Department of Physics, Kinki University, 
Higashi-Osaka, 577-8502 Japan 
} 
\date{\today}


\maketitle 

\begin{abstract}  
In this paper, we study the three-dimensional (3D)
$N_{\rm f}$-flavor CP$^1$ model 
(a set of $N_{\rm f}$ CP$^1$ variables) 
coupled with a dynamical compact U(1) gauge field 
by means of Monte-Carlo simulations.
This model is relevant to 2D $s=1/2$ quantum spin models, 
and has a phase transition line which separates an 
ordered phase of global spin symmetry from a disordered one.
From a gauge theoretical point of view, the ordered phase is a 
Higgs phase whereas the disordered phase is a confinement phase.
We are interested in the gauge dynamics just on the critical line,
in particular, whether a Coulomb-like deconfinement phase is 
realized there. This problem is quite important to clarify 
low-energy excitations in certain class of quantum spin models.
If the gauge dynamics is in the deconfinement phase there, spinons, 
which transform in the fundamental representation of the 
SU($N_{\rm f}$) symmetry, appear as low-energy excitations.
By Monte-Carlo simulations, we found that the ``phase structure" 
on the {\em criticality} strongly depends on the value of $N_{\rm f}$.
For small $N_{\rm f}$, the confinement phase is realized, 
whereas the 
deconfinement phase appears for sufficient large $N_{\rm f}\ge 14$.
This result strongly suggests that compact QED$_3$ is in a 
deconfinement phase for sufficiently large number of flavors
of massless fermions.
\end{abstract} 
]

\setcounter{footnote}{0} 

\section{Introduction}

Some recent experiments 
of strongly-correlated electron systems 
indicate that the usual Fermi liquid theory breaks down in certain
cases, and the low-energy quasi-excitations carrying 
fractional/exotic quantum number appear there.
The fractional quantum Hall effect (FQHE)\cite{CF} 
is a typical example, 
in  which composite fermions appear as relevant excitations.
Another example may be quantum spin models in low spatial
dimensions, which  have been studied quite intensively.
For certain class of $s={1\over 2}$ anti-ferromagnetic 
(AF) spin models
in two dimensions, it is argued that low-energy excitations at a 
quantum phase transition point are spinons\cite{spinon,YAIM}.

For studying the above 
``deconfined critical point" and the quantum phase transition itself,
gauge theory is quite useful. 
Concept of confinement and deconfinement in the gauge theory is suitable
for understanding the change of particle picture happening at deconfined
critical points. 

In the previous paper\cite{YAIM}, we showed that the phase 
transition from the N\'eel state to the dimer state 
in the AF magnet corresponds to a Higgs (deconfinement)
to confinement phase transition in 
the simple CP$^1$ model. 
There we were also 
interested in the gauge dynamics at the critical point.
If the three-dimensional (3D) Coulomb-like phase is realized there as   
a simple loop expansion predicts, quasi-excitations 
are massless spinons.
As the low-energy excitations are magnons (spin waves) 
in the N\'eel state
and  spin-triplet excitations in the dimer state, 
the existence of spinons at the criticality indicates 
breakdown of the traditional
Ginzburg-Landau (GL) theory of phase transition.
This is because the GL theory uses an (a set of)
order parameter 
to describe both a phase transition by its expectation value 
and low-energy excitations by its fluctuations in space and time.

To study the gauge dynamics of a class of spin models 
in a more  general framework, we introduced 
the 3D CP$^1$+U(1) lattice gauge theory in Ref.\cite{TIM}.
The model contains two parameters, the spin stiffness
 $c_1$ and the gauge coupling $c_2$, and describes the
 O(3)(CP$^1$) and O(4) spin models in the specific limits,
($c_2=0$ and $c_2=\infty$, respectively).
However, from the calculation of 
both instanton density and gauge-boson mass, we concluded there 
that the {\em confinement phase} is realized 
on its critical line\cite{reason}.

In this paper, we continue to study the gauge dynamics of  
these spin systems defined in two spatial dimensions
at zero temperature by  generalizing the above 
3D CP$^1$+U(1) model to the 3D multi-flavor 
CP$^1$+U(1)  model.
In particular, we explore the possibility of change of particle
picture on the criticality by controlling the flavor number
$N_{\rm f}$ as an adjustable parameter.

The rest of the paper is organized as follows.
In Sect.2, we explain the model and its relation to the 
AF Heisenberg model.
In Sect.3, results of Monte-Carlo simulations are shown.
We calculated the specific heat, the gauge-boson mass, 
and the instanton density for various values of $N_{\rm f}$, 
and found that the deconfinement phase
is realized on the critical line for sufficiently large $N_{\rm f}$.
Section 4 is devoted for conclusion.

\setcounter{equation}{0}
\section{Multi-flavor CP$^1$ + U(1) model on the 3D lattice}

Let us first define the model on the cubic lattice, and explain its relation
to quantum spin models.
Hereafter we use $x$ as the site index and $\mu=1,2,3$ as 
the direction index.
On each site $x$, we put $N_{\rm f}$-flavor 
CP$^1$ variables $z^\alpha_{x}$, where $\alpha$ is the flavor index and 
takes $\alpha=1, \cdots, N_{\rm f}$.
$z^\alpha_{x}$ is a two-component complex field,
\begin{equation}
z^\alpha_x\equiv \left(\begin{array}{c}
z^\alpha_{x1}\\
z^\alpha_{x2}\end{array}
\right), \,\,\,\,\,\,\,\,\,\, z^\alpha_{x1},z^\alpha_{x2}\in 
\mbox{{\bf C}},
\end{equation}
satisfying the so-called CP$^1$ constraint,
\begin{equation}
\bar{z}_x^\alpha z_x^\alpha= \sum_{a=1,2} |z^\alpha_{xa}|^2=1\ \ 
 {\rm for\ each\ }x\ {\rm and}\  \alpha.
\label{CP1con}
\end{equation}
On each link $(x,x+\mu)$ we put a U(1) gauge variable,
$U_{x\mu}=\exp(i\theta_{x\mu})$ [$\theta_{x\mu}\in (-\pi,+\pi)$].
The action of the model $S$ is given as 
\begin{eqnarray}
S&=&-\frac{c_1}{2}\sum_{x,\mu,a,\alpha}
\Big(\bar{z}^\alpha_{x+\mu,a}U_{x\mu} z^\alpha_{xa} + 
\mbox{H.c.}\Big) \nonumber  \\
&& -\frac{c_2}{2}\sum_{x,\mu<\nu}\Big(\bar{U}_{x\nu}\bar{U}_{x+\nu,\mu}
U_{x+\mu,\nu}U_{x\mu}+\mbox{H.c.}\Big),
\label{model_1}
\end{eqnarray}
where $c_1$ and $c_2$ are real parameters of the model.
It is obvious that the action (\ref{model_1}) has a local U(1) gauge
symmetry as well as SU(2) and SU($N_{\rm f}$) global symmetries;
\begin{eqnarray}
z_{xa}&\rightarrow& z'_{xa}= \exp(i\Lambda_x)z_{xa},\nonumber\\
U_{x\mu}&\rightarrow& U'_{x\mu}= \exp(i\Lambda_{x+\mu})
U_{x\mu}\exp(-i\Lambda_x),
\label{gaugesymmetry}\\
z^\alpha_{xa} &\rightarrow& (z^{\alpha}_{xa})'= 
\sum_{b=1}^2V_{ab} z^\alpha_{xb}, \;\; V \in {\rm SU(2)},
\label{SRS}
\\
z^\alpha_{xa} &\rightarrow& (z^{\alpha}_{xa})'=
 \sum_{\beta=1}^{N_{\rm f}} W^{\alpha\beta} z^\beta_{xa}, \;\;
W \in \mbox{SU($N_{\rm f}$)}.
\label{NfS}
\end{eqnarray}
Hereafter we call the above SU(2)(SU($N_{\rm f}$)) symmetry 
the spin symmetry (flavor symmetry).
The partition function $Z$ is given by
\begin{eqnarray}
Z&=& \int [dU]_{\rm U(1)} [dz]_{\rm CP^1} \exp(-S).
\label{model}
\end{eqnarray}

There are many gauge-invariant quantities
composed of $z_{x}^\alpha$ and $U_{x\mu}$. Among them, 
typical {\it local} combinations are 
a set of $N_{\rm f}$ O(3) (real three-component) 
spins $\mbox{\boldmath $n$}_x^\alpha$,
\begin{eqnarray}
\mbox{\boldmath $n$}_x^\alpha = \bar{z}_x^\alpha 
\mbox{\boldmath $\sigma$} z_x^\alpha,\;
\mbox{\boldmath $n$}_x^\alpha \cdot \mbox{\boldmath $n$}_x^\alpha=1,
\end{eqnarray}
where \mbox{\boldmath $\sigma$}
=$(\sigma_1,\sigma_2,\sigma_3)^{\rm t}$ are
 the Pauli matrices.
For $N_{\rm f}=1$, 
the model reduces in the limit of  $c_2=0$ to 
the O(3) spin model described by the field {\boldmath $n$}$_x$ 
with nearest-neighbor interactions\cite{TIM}.
For  $N_{\rm f}>1$, there are other local 
gauge invariant objects
$\vec{M}^{\alpha\beta}_{x}$, which are
 {\it four-component} O(4) vectors,
\begin{eqnarray}
&&\vec{M}^{\alpha\beta}_{x}=\frac{1}{\sqrt{2}}
(\bar{z}^\alpha_{x}\sigma_1 z_x^\beta,\
\bar{z}^\alpha_{x}\sigma_2 z_x^\beta,\
 \bar{z}^\alpha_{x}\sigma_3 z_x^\beta,\ 
\bar{z}_x^\alpha z_x^\beta)^{\rm t},  \nonumber\\
&& 
\vec{M}^{\alpha\beta *}_{x}\cdot \vec{M}^{\alpha\beta}_{x} =1,\
\vec{M}_x^{\beta\alpha}=\vec{M}_x^{\alpha\beta *}.
\end{eqnarray}
 $\vec{M}^{\alpha\beta}_{x}$ are complex for $\alpha \neq \beta$, 
 while $\vec{M}^{\alpha\alpha}_{x}$ are real. 
In the limit of $c_2 = 0$, one can integrate over
$U_{x\mu}$ link by link to obtain
\begin{eqnarray}
Z_{c_2=0} &=& \int [dz]_{\rm CP^1}\exp\left[\sum_{x,\mu}
\log {\rm I}_0(\gamma_{x\mu})\right], 
\nonumber\\
\gamma_{x\mu}^2&=&\frac{c_1^2}{4}
\sum_{\alpha,\beta=1}^{N_{\rm f}}
\vec{M}^{\alpha\beta}_{x+\mu} \cdot\vec{M}_x^{\beta\alpha}+{\rm H.c.}
\end{eqnarray}
where I$_0(\gamma_{x\mu})$ is the modified Bessel function.
We note that 
$\vec{M}^{\alpha\beta}_{x}$ are {\it not} all independent, 
so one needs to include extra interactions 
associated with the change of variables
from $z_{xa}^\alpha$ to $\vec{M}_x^{\alpha\beta}$ to 
treat them as independent O(4) complex spin vectors.
For finite $c_2$, the model involves nonlocal and/or
nonpolynomial interactions among $\vec{M}_x^{\alpha\beta}$.

On the other hand, in the limit of $c_2 \rightarrow \infty$,
the gauge configuration is restricted to $U_{x\mu}=1$
up to gauge transformations (\ref{gaugesymmetry}).
Then the model reduces to  an ensemble of {\it independent} 
$N_{\rm f}$ O(4) nonlinear sigma models,
\begin{eqnarray}
Z_{c_2=\infty}  &=& \int [dR]_{\rm O(4)}\exp(c_1 
\sum_{x,\mu,\alpha} \vec{R}_{x+\mu}^\alpha\cdot \vec{R}_{x}^\alpha), 
\nonumber\\
{\rm [}dR{\rm ]}_{\rm O(4)}
&=&\prod_{x,\alpha}\prod_{k=1}^4 dR_{xk}^\alpha 
\prod_{x,\alpha}\delta(
\vec{R}_{x}^\alpha\cdot\vec{R}_{x}^\alpha -1),
\nonumber\\
z_{x1}^\alpha&=&R_{x1}^\alpha+iR_{x2}^\alpha,\ 
z_{x2}^\alpha=R_{x3}^\alpha+iR_{x4}^\alpha,
\end{eqnarray}
where $\vec{R}_{x}^{\alpha}$ is a four-component 
real O(4) vector, $\vec{R}_{x}^{\alpha}=(R_{x1}^\alpha,\cdots,
R_{x4}^\alpha)^t$.

\begin{figure}[htbp]
\begin{center}
\leavevmode
\epsfxsize=7cm
\epsffile{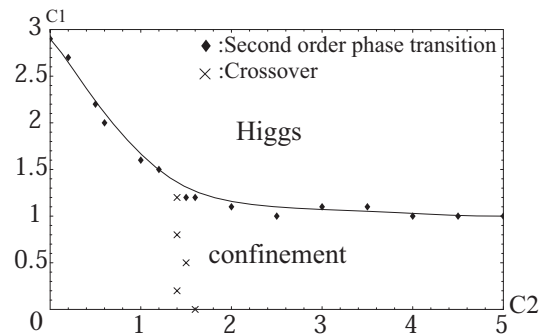}
\caption{Phase structure of the  3D CP$^1+U(1)$ model 
($N_{\rm f}=1$) in the $c_1$-$c_2$ plane
obtained by the measurement of the specific heat\cite{TIM}.
The Higgs and confinement phases correspond to the N\'eel and dimer
state of the quantum spin model, respectively
}
\label{fig.ps}
\end{center}
\end{figure}

In the previous paper\cite{TIM}, 
we investigated the phase structure of the $N_{\rm f}=1$
model and found that there exist two phases, the ordered phase 
of the symmetry (\ref{SRS}) and the disordered phase which are
separated by the second-order transition line $c_{1}=c_{1c}(c_2)$.
(See Fig.\ref{fig.ps}.)
These two phases
correspond to the Higgs and the confinement phases 
in the U(1) gauge dynamics, respectively.
In the ordered phase $c_1 > c_{1c}$, 
there is  a nonvanishing ``spin magnetization"
$\langle \bar{z}_x${\boldmath $\sigma$}$z_x\rangle \neq 0$, and as a result 
the low-energy excitations
are the massless components of $z_{xa}$, which corresponds to 
the spin waves in the AF magnets (see later discussion).
On the other hand, in the disordered phase $c_1 < c_{1c}$, 
the confinement phase is realized,
and the low-energy excitations are the ``spin-triplet" vector field
which is nothing but the composite field, 
$
\mbox{\boldmath $n$}_x = \bar{z}_x\mbox{\boldmath $\sigma$}z_x.
$
Just on the critical line $c_1=c_{1c}$, 
there is no spontaneous symmetry
breaking of the internal spin symmetry and $z_{xa}$ $(a=1,2)$ behave as 
gauge-interacting massless bosons.
Thus one may naturally expect that a 3D Colulomb-like phase with 
a potential $1/r$ may be realized there because 
of the screening effect by the massless bosons $z_{xa}$.
In such a phase on the critical line, the low-energy excitations are
to be  
{\em ``weakly interacting massless spinons"} $z_{xa}$.
With this possibility in mind, 
we studied the gauge dynamics on the critical line, 
and found that {\em the confinement phase is realized there}.
This result means that the CP$^1$ model coupled with 
the dynamical gauge field (\ref{model_1}) belongs to the
same universality class as the $O(3)$ nonlinear $\sigma$ model.

\begin{figure}[bthp]
\begin{center}
\leavevmode
\epsfxsize=7cm
\epsffile{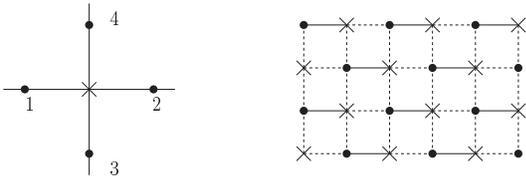}
\caption{2D square lattice; crosses are odd sites, and filled
circles are even sites. Solid bonds indicate that their
exchange couplings are stronger than those on the dotted bonds.}

\label{fig.dimer}
\end{center}
\end{figure}

The $N_{\rm f}=1$  case of the CP$^1$ model (\ref{model_1}) 
is known to be a low-energy effective field theory
of the nonuniform $s={1 \over 2}$ AF Heisenberg model on a square 
lattice\cite{YAIM,IM,aa},
\begin{equation}
H_{\rm AF}=\sum_{x,j}J_{xj}\hat{\mbox{\boldmath $S$}}_x
\cdot\hat{\mbox{\boldmath $S$}}_{x+j}
+\cdots,
\label{AFH}
\end{equation}
where $j$ is the spatial direction
index ($j=1,2$), $\hat{\mbox{\boldmath $S$}}_x$ is the quantum spin operator 
at site $x$, and $J_{xj}$ is the nonuniform exchange coupling.
(See Fig.\ref{fig.dimer}.)
The ellipses in Eq.(\ref{AFH}) represent other multi-spin 
interactions. By varying the couplings $J_{xj}$, the 
{\em ground state} of the Hamiltonian

\begin{figure}[bthp]
\begin{center}
\leavevmode
\epsfxsize=8cm
\epsffile{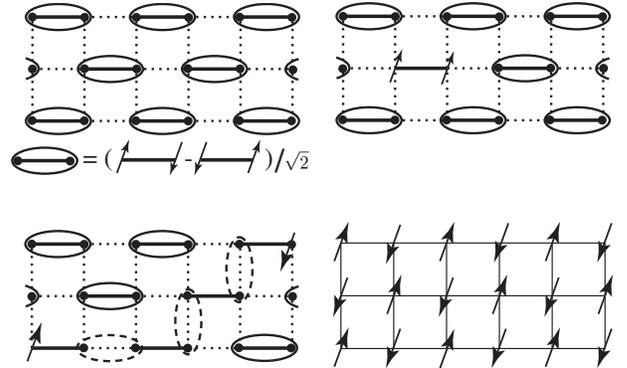}
\caption{Ground states and excitations of the nonuniform AF 
Heisenberg model of Eq.(\ref{AFH}). Each arrow represents one of the 
two  spinon states $z_{x}=(1,0)^t$ and $z_{x}=(0,1)^t$.
Each oval represents a singlet pair of NN spins.
(a) dimer state, (b) spin-triplet excitation in the dimer state,
(c) two-spinon state having the energy proportional to 
their distance (confinement phenomenon),
(d) N\'eel state with an  AF long-range order. }
\label{fig.spin}
\end{center}
\end{figure}

\noindent
(\ref{AFH}) changes from 
the N\'eel state to the dimer state\cite{ND}.
In the N\'eel state, there exists an AF long-range order and 
the low-energy excitations are the spin waves. 
On the other hand, the ground state of the dimer state consists of
spin-singlet pairs on nearest-neighbor (NN) sites, 
and low-energy excitations are the spin-triplets.
(See Fig.\ref{fig.spin}.)

The quantum spin operator $\hat{\mbox{\boldmath $S$}}_x$ is expressed by 
the Schwinger boson, i.e., the CP$^1$ boson operator, as follows;
\begin{equation}
\hat{\mbox{\boldmath $S$}}_x={1\over 2}z^\dagger_x 
\mbox{\boldmath $\sigma$} z_x,
\end{equation}
and the CP$^1$ constraint, $
\sum_a z_{xa}^\dagger z_{xa}|{\rm phys}\rangle=|{\rm phys}\rangle$,
restricts the magnitude of the spin to ${1\over 2}$.
It was shown that the above N\'eel-dimer phase transition 
is nothing but the transition of the 
CP$^1$ model discussed above\cite{YAIM}.
Therefore our investigation on the critical behavior of 
the CP$^1$ model\cite{TIM} indicates that
the {\em quantum} phase transition in the system (\ref{AFH}) 
belongs to the {\em same universality class} with the 
{\em classical} phase transition 
in the 3D $O(3)$ nonlinear $\sigma$ model\cite{senthil}.

The above result supports the traditional idea for quantum phase 
transition that a {\em quantum} 
system in $d$ spatial dimensions belongs to the same 
universality class of a certain {\em classical} system in $d+z$ 
dimensions, 
where $z$ is called dynamical critical exponent\cite{DCE}.
In recent years, however, it has been recognized that the above 
idea of the dynamical exponent breaks down in some cases; 
nontrivial physics appears
at the criticality of {\em quantum} phase transitions.
The spinons, as it was explanied above, 
are a typical example of such interesting possiblity.

In order to see the above interesting phenomenon of quantum phase
transition, we shall extend the model. 
The CP$^N$ model in the 3D continuum 
space-time is certainly such a model, which can be studied by 
the $1/N$ expansion\cite{CPN}.
In the leading order of the $1/N$, a nontrivial infrared fixed point
appears. On this fixed point, i.e., on the critical point, 
a nonlocal term for the gauge field $A_\mu(x)$ like,
\begin{eqnarray}
&& N\int d^3x \int d^3y \sum_{\mu, \nu}F_{\mu\nu}(x)
{1\over |x-y|^2}F_{\mu\nu}(y),\nonumber \\ 
&& F_{\mu\nu}=\partial_\mu A_\nu-\partial_\nu A_\mu,
\label{LA}
\end{eqnarray}
appears in the effective action due to the vacuum polarization 
of the massless $z_x$. 
At long distances, the above term dominates 
the usual Maxwell term which may exist in the original action. 
From (\ref{LA}),
it is straightforward to calculate the potential energy $V(r)$ 
between the two charges
separated by a distance $r$ as $V(r) \propto 1/r$.
Then it is quite interesting to study the CP$^N$ models on the 
lattice for various values of $N$, in particular, to investigate 
the change in their critical behavors.

Below we shall study  the multi-flavor CP$^1$ model 
(\ref{model_1}) numerically instead of the CP$^N$ model.
The reason to choose the multi-flavor
CP$^1$ model is simply a matter of simplicity and shorter
computing time in Monte-Carlo simulations.
In the large-$N$ limit, it is expected that the both models exhibit
similar behavior.\cite{FN}

\setcounter{equation}{0}
\section{Numerical Results}

In this section, we present the results of our numerical
study of the model  on the 3D cubic lattice
of the system size $N=L^3,\ L=8,12,16$ with the
periodic boundary condition for the flavor number 
$N_{\rm f}=1,2,3,4,5,10,14$, and $18$. We measured the internal energy,
the specific heat, the mass of the gauge boson, and the instanton density.
We observed no hysteresis in the internal energy $\langle S \rangle/N$.

\subsection{Specific heat}

We first show the results of the specific heat $C\equiv\langle
(S-\langle S\rangle)^2\rangle/N$ measured 
in order to determine the phase structure.
\begin{figure}[htbp]
\begin{center}
\leavevmode
\epsfxsize=6cm
\epsffile{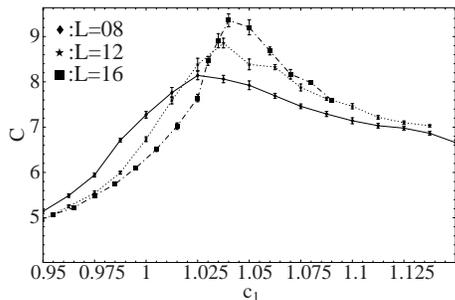}
\caption{Specific heat $C$ vs. $c_1$ at $c_2=2.0$ 
for the $N_{\rm f}=2$ case.
The system size is $8^3, 12^3$ and $16^3$.
$C$ shows typical behavor of the second-order phase transition.
}
\label{sh2.ps}
\end{center}
\end{figure}

\begin{figure}[htbp]
\begin{center}
\leavevmode
\epsfxsize=6cm
\epsffile{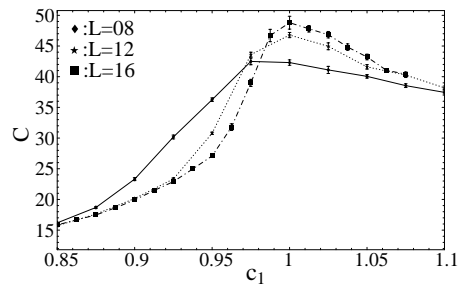}
\caption{Specific heat $C$ vs. $c_1$ at $c_2=2.0$ 
for the $N_{\rm f}=14$ case.
}
\label{sh14.ps}
\end{center}
\end{figure}

\begin{figure}[htbp]
\begin{center}
\leavevmode
\epsfxsize=8cm
\epsffile{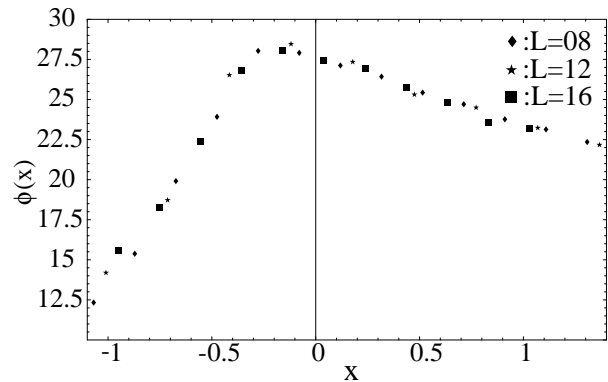}
\caption{Finite-size scaling function $\phi(x)$ of Eq.(\ref{FSSH})
determined by using  $C$ of Fig.\ref{sh14.ps}.
All the data of $L=8,12$ and $16$ are fitted well by the single
function $\phi(x)$.}
\label{FSS.ps}
\end{center}
\end{figure}

In Fig.\ref{sh2.ps} and Fig.\ref{sh14.ps} we present $C$
at the gauge coupling $c_2=2.0$ for $N_{\rm f}=2$ and $14$, respectively.
These results show a typical behavor of 
the second-order phase transition.
To confirm the transition is of second-order, we 
fit these data by the finite-size scaling 
hypothesis (FSSH)\cite{FSS}.
To this end, we introduce 
a parameter $\epsilon\equiv (c_1-c_{1\infty})/
c_{1\infty}$ where $c_{1\infty}$ is the critical 
coupling in the infinite system ($L \rightarrow \infty$).
Then we assume that the correlation length 
at $L \rightarrow \infty$ scales as $\xi \propto \epsilon^{-\nu}$ 
with a critical exponent $\nu$.
We also assume that the maximum of $C$ at $L \rightarrow \infty$,
$C_\infty$ diverges as $C_\infty \propto
\epsilon^{-\sigma}$ with another exponent $\sigma$.
Then FSSH predicts that the specific heat 
$C_L(\epsilon)$ for the system size $L$ scales as 
\begin{equation}
C_L(\epsilon)=L^{\sigma/\nu}\phi(L^{1/\nu}\epsilon), 
\label{FSSH}
\end{equation}
where $\phi(x)$ is the scaling function\cite{FSS}.
The scaling function obtained from the data 
in Fig.\ref{sh14.ps} is shown in Fig.\ref{FSS.ps}.
The parameters are estimated as 
$\nu=1.0, c_{1\infty}=1.01$ and $\sigma=0.20$.
The function $\phi(x)$ is well determined, 
and it is obvious that the FSSH is satisfied quite well.
We investigated the phase structure of the CP$^N$ ($N=2,3,4$) 
models and also $N_{\rm f}=2,3$ cases of the multi-flavor CP$^1$ 
models by calculating the specific heat. 
We conclude that the phase 
structures of both the multi-flavor CP$^1$+U(1) model 
and the CP$^N$+U(1) model are similar to that of  
the CP$^1$+U(1) model shown in Fig.\ref{fig.ps}.

\subsection{Mass of gauge boson}

\begin{figure}[htbp]
\begin{center}
\leavevmode
\epsfxsize=6cm
\epsffile{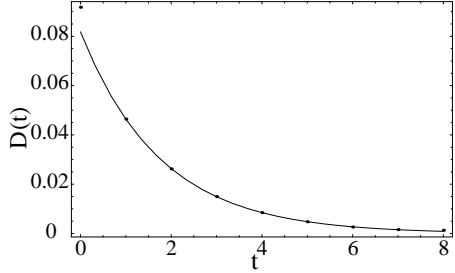}
\caption{Gauge correlation function $D_G(t)$ of Eq.(\ref{CG}) 
for $N_{\rm f}=18$ at $c_1=0.90$ and $c_2=2.0$ with $L=16$.}
\label{DGt.ps}
\end{center}
\end{figure}

Now let us turn to the gauge-boson mass.
We calculate the gauge-invariant gauge-boson mass $M_{\rm G}$ as 
follows\cite{TIM,mass}.
To define $M_{\rm G}$ we first introduce a gauge-invariant operator $O(x)$,
\begin{eqnarray}
O(x)&=&\sum_{\mu,\nu=1,2}
\epsilon_{\mu\nu}\mbox{Im} \; (\bar{U}_{x\nu}\bar{U}_{x+\nu,\mu}
U_{x+\mu,\nu}U_{x\mu})  \nonumber  \\
&=&\sum_{\mu,\nu}
\epsilon_{\mu\nu}\sin (-\theta_{x\nu}-\theta_{x+\nu,\mu}
+\theta_{x+\mu,\nu}+\theta_{x\mu}),
\label{O}
\end{eqnarray}
where $\epsilon_{\mu\nu}$ is the antisymmetric tensor.
Then we intorduce the Fourier transformed field $\tilde{O}(x_3)$ as follows,
\begin{equation}
\tilde{O}(x_3)=\sum_{x_1,x_2}O(x)e^{ip_1x_1+ip_2x_2}.
\label{tO}
\end{equation}
We define the gauge correlation function,
\begin{equation}
D_G(t)={1\over L^3}\sum_{x_3}\Big\langle 
\tilde{O}(x_3)\bar{\tilde{O}}(x_3+t)\Big\rangle.
\label{CG}
\end{equation}
In the continuum, $D_G(t)$ is expected to behave as
\begin{eqnarray}
D_G(t) &=& \int dp_3 {e^{ip_3t}\over \vec{p}^2+M_{\rm G}^2} 
\nonumber \\
&\propto& e^{-\sqrt{p_1^2+p_2^2+M_{\rm G}^2}t}.
\label{CG2}
\end{eqnarray}
Typical behavor of the correlator $D_G(t)$ is shown in 
Fig.\ref{DGt.ps}. We determine $M_{\rm G}$ by fitting the data 
$D_G(t)$ by the exponential form (\ref{CG2}). 
For practical calculations, we set $p_1=p_2=2\pi/L$.

\begin{figure}[htbp]
\begin{center}
\leavevmode
\epsfxsize=8cm
\epsffile{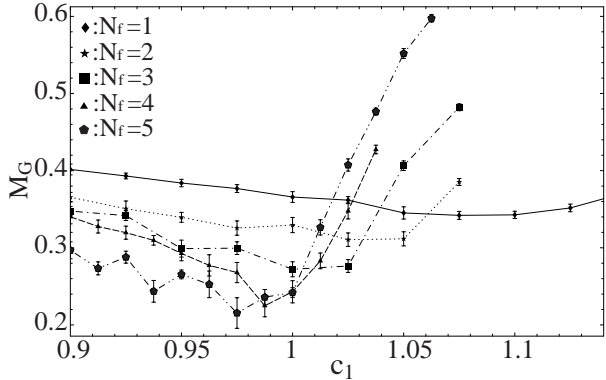}
\caption{Gauge-boson mass $M_{\rm G}$ vs. $c_1$ at $c_2=2.0$
for $N_{\rm f}=1,\cdots,5$. $M_{\rm G}$ do not vanish for these $N_{\rm f}$'s.
}
\label{gbm1.ps}
\end{center}
\end{figure}

\begin{figure}[htbp]
\begin{center}
\leavevmode
\epsfxsize=8.5cm
\epsffile{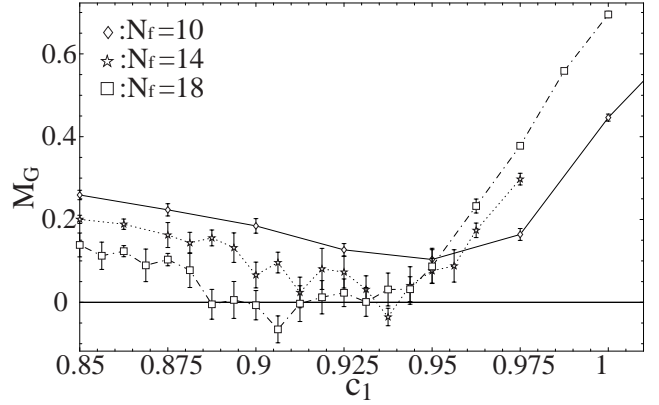}
\caption{Gauge-boson mass $M_{\rm G}$  vs.  $c_1$ at $c_2=2.0$
for $N_{\rm f}=10,14$ and $18$.
The data show that $M_{\rm G}$ vanishes in the critical region 
for $N_{\rm f}=14$ and $18$. The case of $M_{\rm G}<0$ implies
that the square of exponent of Eq.(\ref{CG2}),
$\gamma \equiv M_{\rm G}^2+p_1^2+p_2^2$ is smaller 
than $2(2\pi/L)^2$ and we defined 
$M_{\rm G}\equiv -[2(2\pi/L)^2-\gamma]^{1/2}$.
}
\label{gbm2.ps}
\end{center}
\end{figure}

In Fig.\ref{gbm1.ps} and Fig.\ref{gbm2.ps},
we plot $M_{\rm G}$ for $N_{\rm f}=1\sim 5$ and $N_{\rm f}=10,14,18$, 
respectively.
From the results in Fig.\ref{gbm1.ps}, it is obvious that the 
gauge-boson mass has the minimum in the region close to 
the phase transition point.
The minimum of the value of $M_{\rm G}$ decreases as $N_{\rm f}$ 
increases as expected, but it is still nonvanishing.
In the previous paper\cite{TIM}, we observed similar behavior 
of $M_{\rm G}$ in the CP$^N$+U(1) model for $N=1,2,3,4$.

On the other hand, $M_{\rm G}$ for $N_{\rm f}=10, 14, 18$ 
in Fig.\ref{gbm2.ps} shows that $M_{\rm G}$ vanishes
at the criticality for $N_{\rm f} \ge 14$.
This indicates that a deconfinenemt phase is realized 
on the critical line for large $N_{\rm f}$.
Appearance of the deconfinement phase stems from the shielding effect
by the massless bosons $z^\alpha_x$.
On the critical line, low-energy excitations are 
massless $z^\alpha_x$
and massless gauge boson $\theta_{x\mu}$.
Furthermore, we expect that topological nontrivial excitations, 
i.e., instantons, become irrelevant on the critical line 
due to a large number of the massless $z^\alpha_x$.
(See later discussion.)

\begin{figure}[htbp]
\begin{center}
\leavevmode
\epsfxsize=7cm
\epsffile{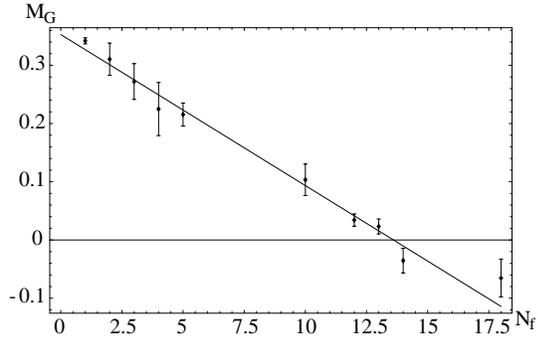}
\caption{The minimum value of gauge-boson mass $M_{\rm G}$ 
for $c_2=2.0$ vs.  the flavor number
$N_{\rm f}$. The straight line interpolates the data
for $N_{\rm f}=1 \sim 5,10$, which intercepts $M_{\rm G}=0$ at
$N_{\rm f} =  13.54$. 
}
\label{fig.mass-nf}
\end{center}
\end{figure}

From the data of Fig.\ref{gbm1.ps} and Fig.\ref{gbm2.ps},
one can locate the minimum value of 
$M_{\rm G}$ along the line $c_2=2.0$ 
for each $N_{\rm f}$. These minima are presented in 
Fig.\ref{fig.mass-nf}. The minimum value of $M_{\rm G}$ seems to
decrease continuously as $N_{\rm f}$ increases.
 By making the linear extrapolation
of the data for $N_{\rm f}=1\sim5$ and 10, we estimate that 
$M_{\rm G}$ starts to vanish at $N_{\rm f} \simeq 13.5$.
This value is regarded as the  critical flavor number at
which the phase on the criticality changes 
continuously from the confinement 
phase to the deconfinement Coulomb-like phase.

\subsection{Instantons}

Instantons play an important role in compact U(1) gauge theories
\cite{inst}.
Their proliferation (condensation) enhances fluctuations of U(1)
gauge field and induces the confinement phase of the gauge dynamics.
In the present 3D case, the instantons are just  the magnetic monopoles
and their condensation puts the system into the ``dual" superconducting
phase. The dual Meissner effect squeezes electric fluxes
one-dimensionally, and as a result a pair of oppositely charged point
particles separated by a distance $r$ have the energy propotional
to $r$, i.e., they are confined.

In order to measure the instanton density, let us define instanton 
charge as in Ref.\cite{instanton}.
The magnetic flux $\Theta_{x,\mu\nu}$
penetrating plaquette $(x,x+\mu,x+\mu+\nu,x+\nu$) is given as,
\begin{eqnarray}
&& \Theta_{x,\mu\nu}\equiv \theta_{x\mu}+\theta_{x+\mu,\nu}
-\theta_{x+\nu,\mu}-\theta_{x\nu}, \nonumber \\
&& \hspace{2cm} (-4\pi<\Theta_{x,\mu\nu}<4\pi).
\label{Theta}
\end{eqnarray}
We decompose $\Theta_{x,\mu\nu}$ into its 
{\it integer} part $2\pi n_{x,\mu\nu}$ ($n_{x,\mu\nu}$ is an integer)
and the remaining part  $\tilde{\Theta}_{x,\mu\nu} \equiv$
 $\Theta_{x,\mu\nu}\;\; (\mbox{mod} \;2\pi$),
\begin{equation}
\Theta_{x,\mu\nu}=2\pi n_{x,\mu\nu}+\tilde{\Theta}_{x,\mu\nu}, \;\;
(-\pi<\tilde{\Theta}_{x,\mu\nu}<\pi).
\end{equation}
Physically, $n_{x,\mu\nu}$ describes the Dirac string. 
The instanton charge $Q_x$ at the cube 
around the dual site $\tilde{x} =x+(\hat{1}+\hat{2}+\hat{3})/2$
is defined as 

\begin{eqnarray}
Q_x&=&
-{1\over 2}\sum_{\mu,\nu,\rho}\epsilon_{\mu\nu\rho}
(n_{x+\mu,\nu\rho}-n_{x,\nu\rho})\nonumber\\
&=&{1\over 4\pi}\sum_{\mu,\nu,\rho}\epsilon_{\mu\nu\rho}
(\tilde{\Theta}_{x+\mu,\nu\rho}-\tilde{\Theta}_{x,\nu\rho}),
\label{instden}
\end{eqnarray}
where $\epsilon_{\mu\nu\rho}$ is the complete antisymmetric tensor.
Then it is obvious that $Q_x$ measures the total flux emanating
from the monopole(instanton) sitting at $\tilde{x}$.
The instanton density $\rho$ is defined as 
\begin{equation}
\rho=\sum_x |Q_x|/N.
\label{rho}
\end{equation}

\begin{figure}[htbp]
\begin{center}
\leavevmode
\epsfxsize=8cm
\epsffile{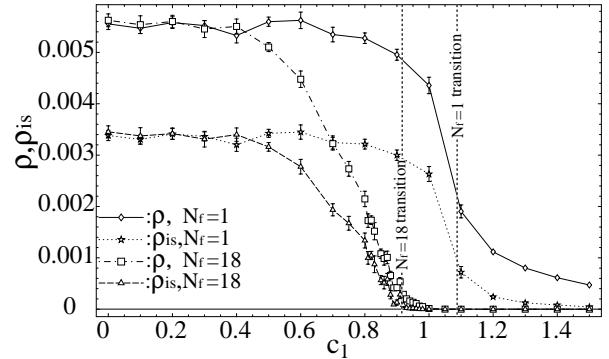}
\caption{Density of instantons $\rho$ and that of isolated instantons
$\rho_{\rm{is}}$ at $c_2=2.0$ vs. $c_1$ for 
$N_{\rm f}=1$ and $N_{\rm f}=18$.
It is obvious that both $\rho$ and $\rho_{\rm{is}}$ for $N_{\rm f}=18$ tend 
to vanish for $c_1 \ge c_{1c}$, whereas $\rho$ and $\rho_{\rm{is}}$ 
for $N_{\rm f}=1$ remain finite at the critical point.
}
\label{inst.ps}
\end{center}
\end{figure}

In Fig.\ref{inst.ps}, we show the instanton density
$\rho$ at $c_2=2.0$ as a function $c_1$ for 
the $N_{\rm f}=1$ and $N_{\rm f}=18$ cases.
As snapshots of instanton configurations in Ref.\cite{TIM} show,
some of instantons form  pairs with anti-instantons
located at NN sites, i.e., instanton-anti-instanton dipoles.
These dipoles are not effective for disordering
the gauge-field dynamics and do not contribute to confinement.
In fact, the confinement phase of the gauge dynamics is nothing
but the plasma phase of the instantons as first shown by 
Polyakov\cite{polyakov}.
On the other hand, the insulating phase of the instantons,
in which almost all instantons form dipoles, is the deconfinement
phase of the gauge dynamics.
Then the density of {\em isolated (single)}
instantons is a physical quantity which monitors whether the system is 
in the (de)confinement phase.
Therefore in Fig.\ref{inst.ps}, we also show the
density of isolated instantons $\rho_{\rm{is}}$,
\begin{equation}
\rho_{\rm{is}}\equiv \rho-2\rho_{\rm{dp}},
\label{is}
\end{equation}
where $\rho_{\rm{dp}}$ is the density of
NN instanton-anti-instanton dipoles defined similarly as in Eq.(\ref{rho})
(The factor $2$ in front of $\rho_{\rm{dp}}$ in Eq.(\ref{is})
comes from the fact that a dipole
is composed of an instanton and an anti-instanton).

In Fig.\ref{inst.ps}, both $\rho$ and $\rho_{\rm{is}}$ almost vanish for 
$c_1>c_{1c}$ in 
the $N_{\rm f}=18$ case in which $M_{\rm G}$ vanishes at the 
critical point.
However, in the $N_{\rm f}=1$ case, there remains a finite instanton 
density at the critical point.
This result and the calculation of $M_{\rm G}$ indicate that 
the Coulomb phase is 
realized on the critical line for $N_{\rm f}\ge 14$.

\begin{figure}[htbp]
\begin{center}
\leavevmode
\epsfxsize=8cm
\epsffile{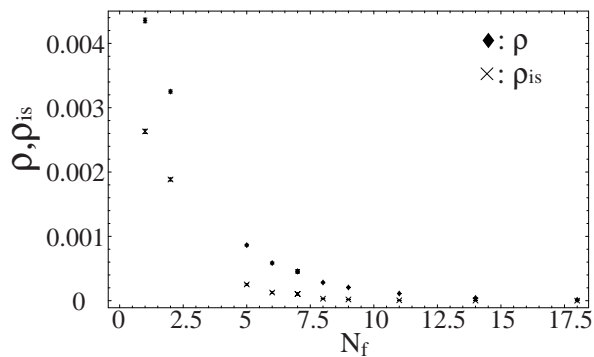}
\caption{Instanton densities $\rho$ and $\rho_{\rm{is}}$
at $c_2=2.0$ and $c_1=1.0$ as a function 
of $N_{\rm f}$.
}
\label{inst2.ps}
\end{center}
\end{figure}

\begin{figure}[htbp]
\begin{center}
\leavevmode
\epsfxsize=8cm
\epsffile{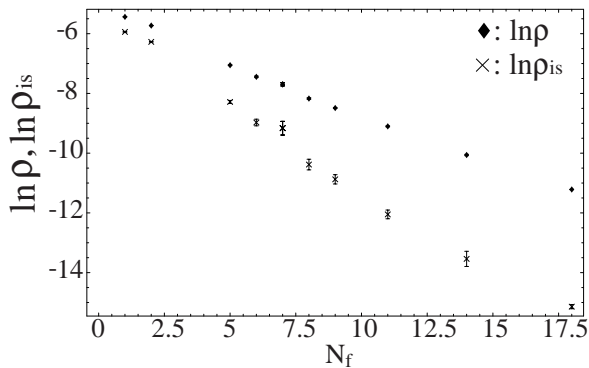}
\caption{Log-plot of instanton densities $\rho$ and $\rho_{\rm{is}}$
at $c_2=2.0$ and $c_1=1.0$ 
as a function of $N_{\rm f}$. It is obvious that $\ln \rho,\
\ln \rho_{\rm{is}} \propto (-N_{\rm f})$.
}
\label{inst3.ps}
\end{center}
\end{figure}

It is interesting to see how the instanton density changes as
a function of the flavor number $N_{\rm f}$.
Results are shown in Figs.\ref{inst2.ps} and \ref{inst3.ps} for 
$c_1=0.1$ and $c_2=2.0$.
From the results,
it is obvious that $\rho$ scales as $\rho \propto e^{-AN_{\rm f}}$
for $N_{\rm f}\ge 5$,
where $A$ is a certain constant.
This means that the main contribution in the effective gauge theory
comes from the vacuum polarization of $z^\alpha_{xa}$
$(\alpha=1, \cdots, N_{\rm f}, \; a=1,2)$.
In fact the effective gauge theory $S_{\rm eff}(U)$
obtained by integrating out the CP$^1$
variables in Eq.(\ref{model}) has the following form,
\begin{equation}
Z=\int [dU]_{\rm U(1)}\; \exp(-S_{\rm eff}),
\label{Seff}
\end{equation}
where
$$
S_{\rm eff}=N_{\rm f}S_z-\frac{c_2}{2}\sum_{x,\mu<\nu}\Big(\bar{U}_{x\nu}
\bar{U}_{x+\nu,\mu}U_{x+\mu,\nu}U_{x\mu}+\mbox{H.c.}\Big),
$$
\begin{equation}
\exp(-S_z)= \int [dz]\exp\Big(\frac{c_1}{2}\sum_{x,\mu,a,\alpha}
(\bar{z}^\alpha_{x+\mu,a}U_{x\mu} z^\alpha_{xa}
+\mbox{H.c.})\Big).
\label{Sz}
\end{equation}
In Eq.(\ref{Sz}), $\int [dz]$ denotes the intergal over
{\em single} CP$^1$ field. 
From the above form of $S_{\rm eff}(U)$, it is expected that 
$S_z(U)$ dominates over the single-plaquette term for sufficiently
large $N_{\rm f}$ and it determines the constant $A$ in the fitting $\rho$
as actually observed in Fig.\ref{inst3.ps}.

\setcounter{equation}{0}
\section{Conclusion}

In this paper we studied the multi-flavor CP$^1$ model in three
dimensions by Monte-Carlo simulations.
In particular, we are interested in the gauge dynamics on the 
critical line which separates the Higgs (N\'eel) and 
confinement (dimer) phases.
On the critical line, ``spinons" $z_x^\alpha$ are massless.
Their fluctuations shield the confining gauge force
at least partly. If the number of these spinons is sufficiently
large, the confining forces may be completely shielded by them
and the deconfining (Coulomb-like) force may appear instead.
By calculating the gauge-boson mass and the instanton density, we 
found that the Coulomb-like deconfinement phase is 
actually realized for $N_{\rm f}\ge 14$.
The low-energy excitations on the critical line 
are the massless ``spinons" $z_x^\alpha$ and massless gauge boson.
Similar deconfinement phase is expected to appear on the critical points 
of the large-$N$ solution of the CP$^N$ model. 

As far as the shielding phenomenon is concerned, massless 
fermions give a similar effect as massless bosons.
Thus the present result indicates that the 
parity-preserving QED$_3$ with massless four-component-spinor 
fermions should have a deconfinement phase for sufficiently large 
flavor number of fermions, as long as the 
chiral symmetry is {\em not} 
spontaneously broken to avoid the generation of the dynamical mass. 
For example, in perturbation theory, 
gauge-interacting fermions generate the nonlocal 
terms like Eq.(\ref{LA}).
Recently,  3D U(1) gauge theories coupled with gapless 
matter fields have been studied quite intensively, 
in particluar, to answer the question whether 
a confinement-deconfinement phase transition takes 
place\cite{u1}.
The results of the present paper are in agreement 
with those obtained in these works.

One may wonder how the results in this paper are applied to the 
dynamics of realistic quantum spin models. 
The corresponding quantum model for the 
$N_{\rm f}$-flavor CP$^1$ model is the 
SU(2)$\times$ SU($N_{\rm f}$) ``spin" AF magnets, whereas for the 
CP$^N$ model that is SU($N+1$) ``spin" AF magnets.
Unfortunately, as far as we know, there are no materials which have
the above internal quantum degrees of freedom.
However, as we explained in the introduction and also in the above, 
study of the strongly-correlated
electron systems often reduces to the study of 
certain gauge models of 
gapless matter fields.
The results in the present paper should give 
an important insight into the  phase structures of these gauge 
systems and the related strongly-correlated electron systems.


\end{document}